\documentclass[preprint,proceedings]{rmaa}

\SetYear{2002}
\SetConfTitle{Galaxy Evolution: Theory and Observations}

\title{A Model for the Formation and Evolution of Cosmological Halos}

\author{
  Marcelo A. Alvarez,\altaffilmark{1,2}
  Paul R. Shapiro,\altaffilmark{1}
  and Hugo Martel\altaffilmark{1}}

\altaffiltext{1}{Department of Astronomy, University of Texas at Austin,
	Austin, TX 78712}
\altaffiltext{2}{DOE Computational Science Graduate Fellow}

\suppressfulladdresses

\listofauthors{
  M. A. Alvarez
  P. R. Shapiro
  H. Martel
}
\indexauthor{
  Alvarez, M. A.
  Shapiro, P. R.
  Martel, H.
}
\shortauthor{Alvarez, Shapiro, \& Martel}
\shorttitle{A Model for the Formation and Evolution of Cosmological Haloes}

\addkeyword{Cosmology: Theory}
\addkeyword{Dark Matter}
\addkeyword{Galaxies: Formation}
\addkeyword{Galaxies: Kinematics and Dynamics}
\addkeyword{Large-scale Structure of Universe}

\begin{document}
\maketitle 

\section{Introduction}
Adaptive SPH and N-body simulations were carried out to study the collapse
and evolution of dark matter halos that result from the gravitational
instability and fragmentation of cosmological pancakes. Such halos
resemble those formed by hierarchical clustering from realistic initial
conditions in a CDM universe and, therefore, serve as a convenient
test-bed model for studying halo dynamics. Our halos are in approximate 
virial equilibrium and roughly isothermal, as in CDM simulations.  Their 
density profiles agree quite well with the fit to N-body results for CDM 
halos by Navarro, Frenk, \& White (1997; NFW). 

This test-bed model enables us to study the evolution of individual halos.
The masses of our halos evolve in three stages: an initial
collapse involving rapid mass assembly, an intermediate stage
of continuous infall, and a final stage in which infall tapers off
as a result of finite mass supply. In the intermediate stage, halo mass 
grows at the rate expected for self-similar spherical infall, with $M(a)
\propto a$.  After the end of initial collapse at ($a\equiv a_0$), 
the concentration parameter grows linearly with
the cosmic scale factor $a$, $c(a)\simeq 4(a/a_0)$. The virial
ratio $2T/|W|$ just after virialization is about 1.35, a value close to
that of the $N$-body results for CDM halos, as predicted by the truncated
isothermal sphere model (TIS) (Shapiro, Iliev, \& Raga 1999) and consistent 
with the value expected for a
virialized halo in which mass infall contributes an effective surface
pressure.  Thereafter, the virial ratio evolves towards the value expected
for an isolated halo, $2T/|W|\simeq 1$, as the mass infall rate declines.  
This mass accretion history and evolution of concentration parameter are
very similar to those reported recently in $N$-body simulations of CDM
analyzed by following the evolution of individual halos.  We therefore
conclude that the fundamental properties of halo collapse, virialization,
structure, and evolution are generic to the formation of cosmological
halos by gravitational instability and are not limited to hierarchical
collapse scenarios or even to Gaussian-random-noise initial conditions.

\begin{figure}[!t]
\begin{center}
  \includegraphics[width=2.5in]{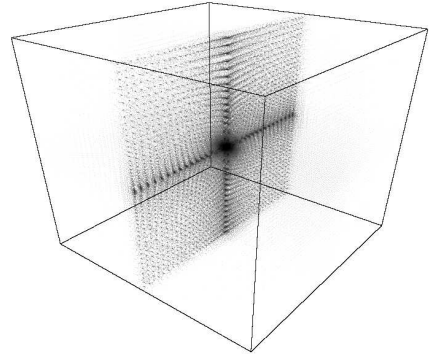}
\end{center}
  \caption{ 
Dark matter particles at $a/a_c=3$.
}
  \label{fig:simple}
\end{figure}

\section{Halo Formation via Pancake Instability}
{\bf Test-bed Model:}  Cosmological pancakes -- modelled as single
plane-wave density fluctuations -- are gravitationally unstable; density
perturbations transverse to the direction of pancake collapse cause the
pancake to fragment (Valinia {\em et al.} 1997).  When a
pancake is perturbed by two transverse density modes with wavevectors in
the plane of pancake collapse, a quasi-spherical halo forms at the
intersection of two filaments in the pancake plane.  This halo closely
resembles those formed by hierarchical clustering from initial conditions
in a CDM universe.

{\bf ASPH/P3M Simulations:} Two simulations were run of the formation of a
dark matter halo by pancake instability for use as a test-bed
model for halo formation, with $64^3$ DM particles, both with and
without $64^3$ gas particles.  The primary pancake collapses (i.e. first
forms accretion shocks and caustics) at a scale factor $a_c$.
By $a/a_c=3$, a well-defined pancake-filament-halo structure 
appears (Fig. 1).

\begin{figure}[!t]
\begin{center}
  \includegraphics[width=2.7in]{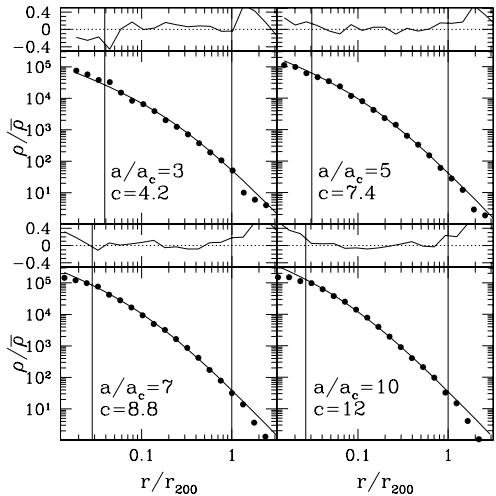}
\end{center}
\caption{ 
Density profiles: spherically-averaged simulation of DM no gas (dots),
best-fit NFW profiles (solid curves), at $a/a_c =$ 3, 5, 7, 10; fractional
deviations from NFW, $(\rho_{\rm NFW}-\rho)/\rho_{\rm NFW}$, above.
}
  \label{fig:simple}
\end{figure}

\begin{figure}[!b]
\begin{center}
  \includegraphics[width=3.0in]{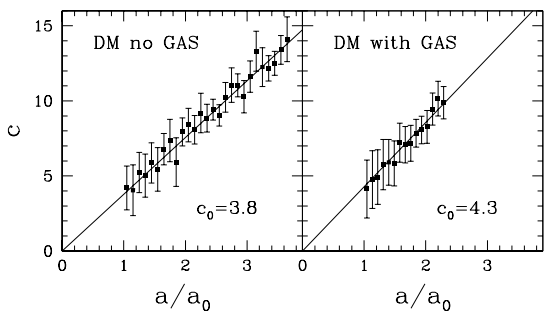}
\end{center}
  \caption{ Concentration parameter vs. scale factor.}
  \label{fig:simple}
\end{figure}

\section{Halo Profiles \& Equilibrium}
{\bf Density Profile:} By $a/a_c=3$, the DM halo (simulated with and 
without gas) has a spherically-averaged density profile close to the 
fit by NFW to N-body simulations of halos formed by hierarchical clustering 
in CDM (Fig. 2).

{\bf Jeans Equilibrium \& Anisotropy:} After $a/a_c=3$, the halo is close to
equilibrium, according to the Jeans equation in spherical symmetry, 
and is close to isothermal.  Hereafter, we shall consider 
the halo to form at $a_0\equiv 3a_c$.
The velocity distribution is somewhat more anisotropic than found in
simulations of CDM, $0.2<\beta<0.8$, whereas the CDM simulations give
$0.0<\beta_{CDM}<0.6$ , where $\beta=1- \langle v_t^2\rangle /(2\langle
v_r^2\rangle)$.

\section{Evolution}
{\bf Concentration Parameter:} The concentration parameter of
the best-fitting NFW density profile at each epoch evolves linearly with
scale factor (Fig. 3).  For $a>a_0$, after the halo formation epoch,
we find $c\simeq 4(a/a_0)$, almost identical to that reported by
Wechsler {\em et al.} (2002) for N-body simulations of CDM halos.

{\bf Mass Growth Rate:} For $2<a/a_c<3$, $M_{200}$ (mass within $r_{200}$) 
grows rapidly, while for
$3<a/a_c<7$, $M_{200}\propto a$, consistent with self-similar spherical
infall (Bertschinger 1985).  For $a/a_c>7$, growth flattens due to finite
mass supply.  This mass history closely resembles that for CDM halos
found by Wechsler {\em et al.} (2002) (Fig. 4). 
 
{\bf Self-Similar Infall:} The radial velocity profile, mass, and
radius are consistent with self-similar infall for $3<a/a_c<7$, with
$\lambda_{200}/\lambda_c\simeq 0.8$, where $\lambda_{200} =
r_{200}/r_{ta}$, $r_{ta}$ is the time-varying turnaround radius, and
$\lambda_c$ is the radius of the outermost caustic in the self-similar
solution.

{\bf Virial Ratio:} The virial ratio $2T/|W|$ just after virialization is
$\sim 1.35$, close to that of the N-body results for CDM halos, as
predicted by the TIS model and consistent with the value 
expected for a virialized halo
in which mass infall contributes an effective surface pressure.
Thereafter, the virial ratio evolves towards the value expected for an
isolated halo, $2T/|W|\sim 1$, as the mass infall rate declines.

\begin{figure}[!t]
\begin{center}
  \includegraphics[width=2.5in]{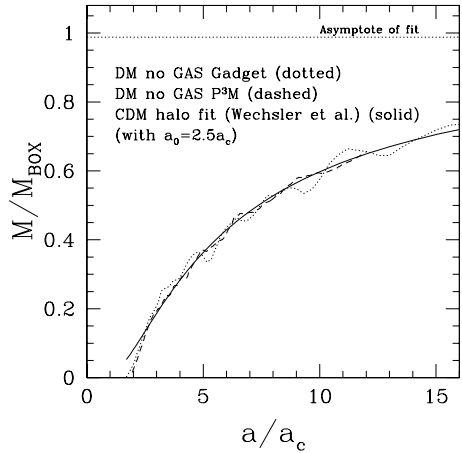}
\end{center}
\caption{ Halo mass vs. scale factor.  ``Gadget'' curve used code of Springel 
{\em et al.} (2001) to simulate same problem.
}
  \label{fig:simple}
\end{figure}
\section*{Acknowledgments}
This work was supported in part by grants NASA ATP NAG5-10825 and
NAG5-10826 and Texas Advanced Research Program 3658-0624-1999.

\end{document}